\documentclass[pra,epsfig,amssymb]{revtex4}

\usepackage{graphicx}

\begin{document}
\title{
How to fix a broken symmetry: Quantum dynamics
of symmetry restoration in a ferromagnetic Bose-Einstein condensate
}
\begin{abstract}
We discuss the dynamics of a quantum phase transition in a spin-1
Bose-Einstein condensate when it is driven from the magnetized broken-symmetry 
phase to the  unmagnetized  ``symmetric'' polar phase. 
We determine  where the condensate goes out of equilibrium 
as it approaches the critical point, and compute the condensate 
magnetization at the critical point. This is done  within a quantum 
Kibble-Zurek scheme traditionally employed in the context of symmetry-breaking
quantum phase transitions.
Then we study the influence of the nonequilibrium  dynamics near 
a critical point on the condensate magnetization.
In particular, when the quench stops at the critical point,
nonlinear oscillations of magnetization occur. They are  characterized by 
a period and an amplitude that are inversely proportional.
If we keep driving the condensate far away from the critical
point through the unmagnetized ``symmetric'' polar phase, 
the amplitude of magnetization oscillations slowly decreases
reaching a non-zero asymptotic value. That process is described by 
the  equation that can be mapped onto the classical mechanical problem
of a particle moving under the influence of harmonic and ``anti-friction''
forces whose interplay  leads to surprisingly simple fixed-amplitude oscillations. 
We  obtain several scaling results relating the condensate
magnetization to the quench rate, and verify numerically all analytical predictions.
\end{abstract}
\author{Bogdan Damski and Wojciech H. Zurek}
\affiliation{
Theory Division, Los Alamos National Laboratory, MS-B213, Los Alamos, NM 87545, USA
}
\maketitle

\section{Introduction}
\label{sec1}

Any symmetry breaking phase transition can be traversed in two
opposite directions. The usual way is to start in the symmetric 
phase, and move into the phase with a broken symmetry. When this process
happens on a finite timescale, critical slowing down will generally 
lead to the random local choices of the  broken-symmetry vacua. This in turn can 
result in the formation of topological defects \cite{kibble,zurek}.
They will appear with the density set by the size of the regions that 
choose the same broken-symmetry vacuum. That size -- and, hence, the
density of the resulting non-equilibrium structures -- can be deduced 
from the critical scalings of the relaxation time and the 
healing length.

The obvious question that has not been addressed 
to date concerns traversing the transition 
from the phase in which the symmetry is broken to where it is restored. Such transition  
can be also induced on a finite timescale, which will again bring 
into the discussion  the critical scalings. Now, however, they will determine 
remaining excitation of the system rather than the size of  broken-symmetry regions.

We study here  the dynamics of quantum phase transitions  \cite{dorner,bodzio,spiny,polkovnikov}
rather than their classical (thermodynamical) counterparts 
that have received extensive  
theoretical \cite{kibble,zurek} and experimental
\cite{eksperymenty,anderson} attention (see \cite{kibble_today}
for a recent review). More precisely, 
we are interested in  the dynamics of a ferromagnetic Bose-Einstein
condensate (BEC) composed of spin-1 atoms \cite{ferro_ref}. Such a 
condensate is studied experimentally by several 
groups \cite{nature_kurn,ferro_exp_jacobi}. Similar issues may 
also arise in an even more complex case of higher spin condensates \cite{pfau}.
The research on symmetry breaking 
in the spin-1 BEC has been the subject of the seminal paper of  
the Berkeley group \cite{nature_kurn}.  There the condensate was
rapidly driven  from the polar (symmetric) to the ferromagnetic (broken symmetry)
phase by the decrease of a magnetic field. Formation of topological defects 
(e.g., vortices) was
observed. This work has triggered several  theoretical 
investigations \cite{ueda_numerics3D,austen,uwe,ferro_bodzio,ueda_1_6,girvin}. 
In particular, vortex density after a slow quench was  found through the 
extensive numerical simulations
by Saito,  Kawaguchi and  Ueda \cite{ueda_1_6} to follow the scaling 
result obtained from the quantum Kibble-Zurek mechanism  \cite{ferro_bodzio}
and the studies of the magnetization correlation functions \cite{austen,ueda_1_6}.

Below we consider  dynamics of a ferromagnetic condensate 
driven through a transition ``in the opposite direction'': from the
broken-symmetry phase to the polar phase.
As in \cite{ferro_bodzio}, we are interested in  slow transitions 
that exhibit adiabatic-impulse behavior so that the condensate 
goes  out of equilibrium {\it close} to the critical point.
In this case the condensate excitation in the polar phase 
reflects the scalings of the critical 
regime. This scenario is supported by the following {\it general} discussion. Suppose
that we change some dimensionless parameter $q$ to 
drive the system towards  the critical point at $q_c$. 
Close to the phase boundary, the gap 
$\Delta$ in the excitation spectrum behaves as 
$\Delta_0(q_c-q)^{z\nu}$, where
$z$ and $\nu$ are critical exponents \cite{sachdev}, $q<q_c$, and $\Delta_0$ is
a constant with dimension of energy. 
As long as the system is driven far enough from the critical point, 
the gap is large and the system excitation does not occur. The relevant 
``reaction time'' for the system  is given by
\begin{equation}
\label{111}
\frac{\hbar}{\Delta}=\tau.
\end{equation}
The system  undergoes adiabatic evolution when $\tau$ is small 
compared to the timescale on which changes occur in  the Hamiltonian. 
That timescale, in turn, is characterized by how fast the ``instantaneous''
excitation gap changes, or in other words by how fast the system is driven. 
It is simply given by 
\begin{equation}
\label{222}
\frac{\Delta}{\left|\frac{d\Delta}{dt}\right|},
\end{equation}
having a proper dimension of time.
Comparison of the  timescales (\ref{111}) and (\ref{222}),  
$$
\frac{\hbar}{\Delta}= \frac{\Delta}{\left|\frac{d\Delta}{dt}\right|},
$$
gives the location of
the border between adiabatic -- non-adiabatic behavior  at some
$q=q_c-\hat q$. 
Assuming, that $\frac{d}{dt}q(t)=\tau_Q^{-1}$ near a critical point
($\tau_Q^{-1}$ is the speed of transition between the phases) we get that 
\begin{equation}
\label{lambda_hat}
\hat q= \left(\frac{z\nu\hbar}{\Delta_0}\right)^\frac{1}{1+z\nu}\frac{1}{\tau_Q^{1/(1+z\nu)}}. 
\end{equation}
The qualitative picture of system dynamics is as follows. For
$q_c-q(t)\gtrsim\hat q$ the system undergoes adiabatic evolution following 
instantaneous changes to its Hamiltonian. When $q_c-q(t)\lesssim\hat q$ 
the evolution ceases to be adiabatic and to a first
approximation one can expect the impulse stage, where the state of the system
does not change -- remains ``frozen''. As a result,  system properties near  the
critical point are determined by its ground state at $q=q_c-\hat q$, where $\hat q$ is 
computed from the quench time $\tau_Q$ and 
the  product of the critical exponents: Eq. (\ref{lambda_hat}).
This approach was already  successfully used to study the  Landau-Zener dynamics \cite{bodzio},
the dynamics of the quantum Ising model \cite{dorner}, and the ferromagnetic 
condensate dynamics during the symmetry-breaking transitions from the
polar to the broken-symmetry phase \cite{ferro_bodzio}.

We begin our study in the next section, where we discuss the basics of a
quantum phase transition
in the mean-field model that we consider. Section \ref{sec3} presents 
how the condensate gets excited when driven through the broken-symmetry phase.
Section \ref{sec4} is devoted to studies of magnetization oscillations 
after  quench. We  point out the observables that contain  the information 
about condensate excitation at the phase boundary. Section \ref{sec5} determines  
the range of applicability of the Single Mode Approximation (SMA) used in 
Sections \ref{sec3} and \ref{sec4}. Finally, Section \ref{sec6} provides the 
summary of the paper.

\section{The model}
\label{sec2}

The energy of the ferromagnetic condensate placed in an external, homogeneous,
magnetic field aligned in the $z$ direction  is given in the mean-field approximation as 
\begin{equation}
\label{eeEE}
{\cal E}[\Psi]= \int d{\bf r} \ \frac{\hbar^2}{2M}|\vec{\nabla}\Psi|^2
       + \frac{c_0}{2}\left(\Psi^\dag\Psi\right)^2 
       + Q\langle\Psi|F_z^2|\Psi\rangle+  
       \frac{c_1}{2}\sum_{\alpha=x,y,z}\langle\Psi|F_\alpha|\Psi\rangle^2,
\end{equation}
where $F_\alpha$ is the  standard spin-1 matrix, and 
$c_0>0$ and $c_1<0$ provide the strength of
spin independent and spin-dependent interactions, respectively:
$$
c_0=\frac{4\pi\hbar^2}{M}\frac{a_0+2a_2}{3}, \ \ c_1=\frac{4\pi\hbar^2}{M}\frac{a_2-a_0}{3},
$$
with $a_S$ being the s-wave scattering length in the total spin $S$
channel, $M$ is the atom mass, while $Q$ (proportional to  square of
magnetic field imposed on the condensate) is the prefactor of the 
quadratic Zeeman shift resulting from atom-magnetic field  interactions. 
The linear Zeeman term is skipped in Eq. (\ref{eeEE}) because it
has a trivial effect on condensate dynamics. Indeed, it 
only rotates the condensate magnetization around the $z$ axis 
with the Larmor frequency. This rotation, of course, can be removed from theoretical
studies by going to a reference frame rotating with the Larmor frequency 
(Sec. II.A of Ref. \cite{ueda_numerics3D}).

The wave function has three 
condensate components, $\psi_m$,  corresponding to $m=0,\pm1$ projections of
spin-1 onto the magnetic field:
\begin{equation}
\label{psi}
\Psi= 
\left(
\begin{array}{l}
\psi_1\\
\psi_0\\
\psi_{-1}
\end{array}
\right), \ \int d{\bf r}\Psi^\dag\Psi= N,
\end{equation}
where $N$ is the total number of atoms. In subsequent calculations we will use
the phases $\chi_m$ defined as: $\psi_m=|\psi_m|\exp(i\chi_m)$.

The simplest meaningful approach  to the dynamics of a ferromagnetic 
Bose-Einstein condensate is provided by the Single Mode Approximation
where the translationally invariant mean-field approach is 
used. This simplification, while obviously inapplicable for
symmetry-breaking phase transitions where the orientation of magnetization 
is explicitly position-dependent \cite{ferro_bodzio,ueda_1_6}, can be successfully
used for the transitions considered here where the system starts from a
uniform broken-symmetry ground state (GS). The influence of 
inhomogeneities  will be illustrated in
Sec. \ref{sec5} where we will compare simulations done with  the SMA to full 
mean-field calculations.
In the SMA both the gradient term and the
density-density interaction term $\sim c_0$ are irrelevant and so are skipped 
in Secs. \ref{sec2}-\ref{sec4}. These terms, however, reappear in  the numerical 
calculations in Sec. \ref{sec5}.

As the ferromagnetic condensate is described in the SMA by  translationally 
invariant spinor (\ref{psi}), it is instructive to parametrize 
the coupling constant $Q$ from (\ref{eeEE}) by
$$
Q= q n|c_1|,
$$
where $n=\Psi^\dag\Psi$ is the atom density, and $q$ is a
dimensionless parameter
proportional to the square of the  magnetic field. This parameter 
will be changed to drive the condensate from one quantum phase to another.

The system longitudinal ($f_z$) and transverse ($f_x$ and $f_y$) magnetizations
read
$$
f_\alpha= \langle\Psi|F_\alpha|\Psi\rangle, \ \alpha= x,y,z.
$$
We consider the
$f_z=0$ case, i.e., 
$|\psi_1|=|\psi_{-1}|$. This condition is dynamically conserved during all the
evolutions considered in Secs. \ref{sec3} and \ref{sec4}
because $\frac{d}{dt}\int d{\bf r}f_z({\bf r},t)=0$,
which in the SMA reduces to $\frac{d}{dt}f_z(t)=0$. The
$f_{x,y}$ components are conveniently combined to a complex transverse magnetization  
\begin{equation}
\label{fT}
f_T = f_x+if_y = \sqrt{2}(\psi_1^*\psi_0+\psi_{-1}\psi_0^*) =
2\sqrt{2}|\psi_1||\psi_0|\cos\left(\chi_0-\frac{\chi_1+\chi_{-1}}{2}\right)
e^{i\phi}, \ \ \phi=\frac{\chi_{-1}-\chi_1}{2}.
\end{equation}

The quantum phases of the ferromagnetic condensate were recently
discussed in \cite{ueda_spectrum}. 
The condensate is in the polar phase when $q>2$. There the ground state 
is given by
$$
\Psi=\sqrt{n}\left(
\begin{array}{c}
0\\
1\\
0
\end{array}
\right),
$$
so that $f_x=f_y=0$. For $0\le q<2$ the system is in the broken-symmetry
phase where the GS spinor reads
\begin{equation}
\label{bsym}
\Psi=\sqrt{n}
\left(
\begin{array}{c}
\sqrt{\frac{1}{4}-\frac{q}{8}}e^{i\chi_1}\\
\sqrt{\frac{1}{2}+\frac{q}{4}}e^{i(\chi_1+\chi_{-1})/2}\\
\sqrt{\frac{1}{4}-\frac{q}{8}}e^{i\chi_{-1}}\\
\end{array}
\right),
\end{equation}
and 
\begin{equation}
\label{fT_}
f_T = n\sqrt{1-\frac{q^2}{4}}e^{i\phi}.
\end{equation}

The system dynamics in the SMA is governed by \cite{jednostki}
\begin{eqnarray}
\label{3gp}
i\hbar\dot{\psi}_1&=& c_1|\psi_0|^2\psi_1 + c_1\psi_0^2\psi_1^*e^{-2i\phi}+
n|c_1| q\psi_1 \nonumber \\
i\hbar\dot{\psi}_0&=& 2c_1|\psi_1|^2\psi_0+ 2c_1\psi_1^2\psi_0^*e^{2i\phi}\nonumber \\
\dot{\phi}&=& 0,
\end{eqnarray}
where $d/dt$ is denoted with a ``dot'' and we have used the identity 
$|\psi_1|=|\psi_{-1}|$ to arrive at Eqs. (\ref{3gp}). The initial condition for
time evolutions in Secs. \ref{sec3} and \ref{sec4}
is provided by the spinor state (\ref{bsym}) taken at $q=0$.
The last equation of (\ref{3gp}) states that the orientation of the transverse
magnetization on the $(x,y)$ plane is conserved throughout the
evolution and so it is fixed by the initial conditions. This constraint 
prohibits full restoration of the symmetry in the state evolved from
the broken-symmetry phase to the symmetric polar phase. In particular, 
$$
f_x(t)\sin(\phi)= f_y(t)\cos(\phi).
$$

We will discuss below the dynamics driven by the variation of $q$
that can be achieved with a proper manipulation of the magnetic field imposed
on the condensate.
A linear increase of the magnetic field strength  in time is well within 
the experimental capabilities \cite{nature_kurn}. Due to the quadratic Zeeman
coupling, it  results  in 
\begin{equation}
\label{q_od_t}
q(t)= \frac{1}{8}\left(\frac{t}{\tau_Q}\right)^2.
\end{equation}
This is the time dependence we will assume in all subsequent calculations.
Close to the phase transition point, where $t=4\tau_Q-\delta t$, we have 
$$
q(\delta t)\approx 2-\frac{\delta t}{\tau_Q}.
$$
For the  transitions considered here the system will cease to adiabatically 
follow the ramp up of $q$ near a critical point where the above
linearization holds: the familiar time-dependence from the Kibble-Zurek theory
emerges near the phase boundary with $\tau_Q^{-1}$ providing the quench rate
\cite{liniowe}, or in other words, the speed in the parameter space 
of driving the system through the critical point.

\section{System excitation}
\label{sec3}

\begin{figure}[t]
\includegraphics[width=0.5\columnwidth,clip=true]{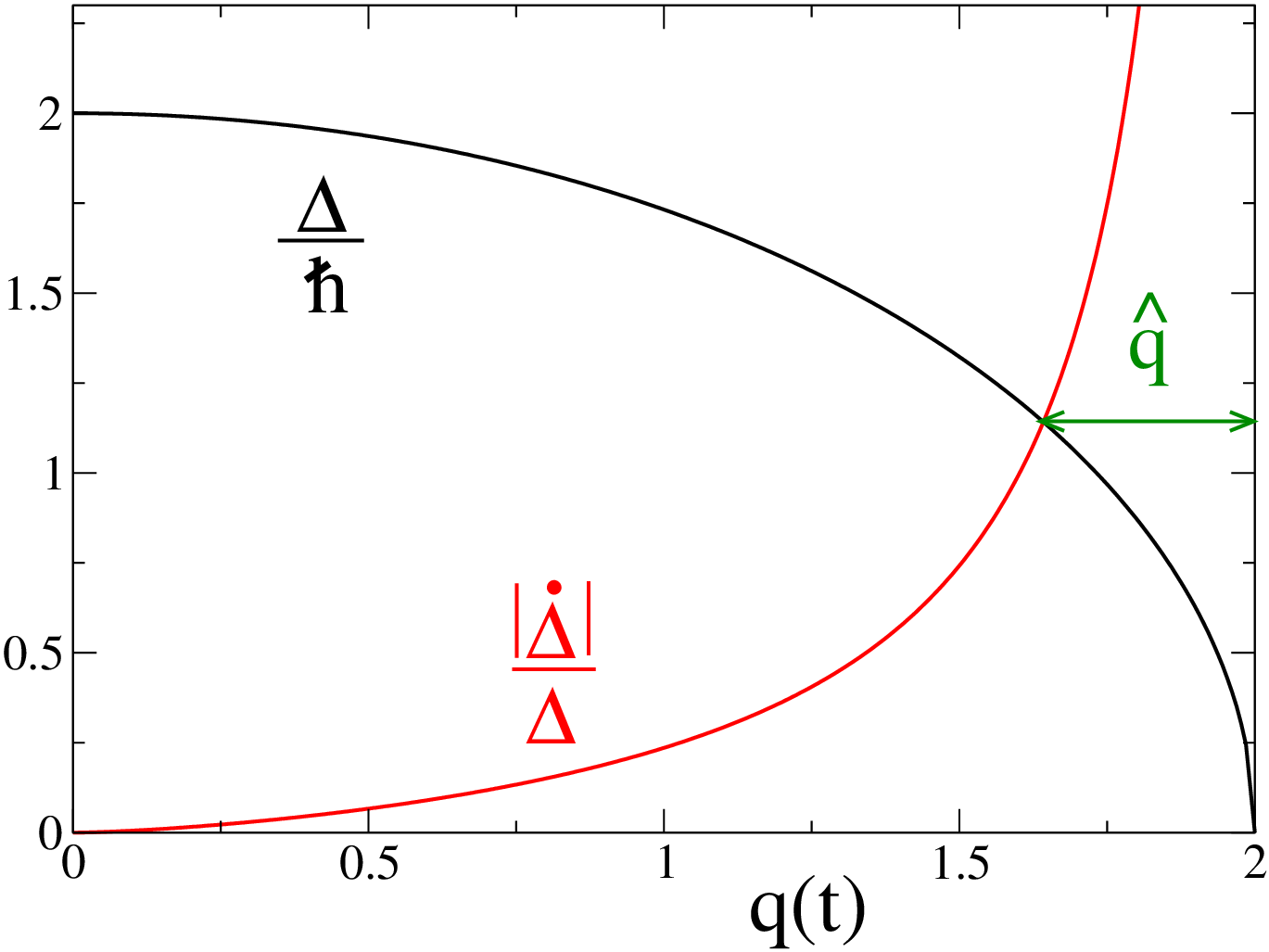}
\caption{(color) Illustration of Eq. (\ref{gap_equation}) in  units of
         $\tau_0^{-1}$  (\ref{qhat}) on the vertical axis. 
	 The plot is for $\tau_Q/\tau_0=1$. 
	}
\label{gap_eq_new}
\end{figure}

For   small perturbations around the GS of the broken-symmetry phase  
one  finds  three Bogolubov modes as in \cite{ueda_spectrum}:
two gapless modes and one gapped.
We do not consider analytically the dynamics induced by the 
excitation of the gapless modes. Instead, we focus on the gapped mode
and show that its excitation is responsible for 
changes of the magnitude of the 
system transverse magnetization during non-equilibrium dynamics. 
This  mode in the long wavelength limit, obviously relevant for the
SMA, has the energy gap \cite{ueda_spectrum}
\begin{equation}
\label{ddd}
\Delta= 2n|c_1|\sqrt{1-q^2/4}.
\end{equation}
As we drive the system to the critical point changing  $q(t)$ 
we expect qualitatively the following dynamics. 
First, for $q(t)$ small enough the evolution
is adiabatic with respect to the gapped mode -- far enough 
from the critical point the 
energy cost of exciting the gapped mode is too large. However,  as the critical
point is approached, the gap becomes too small to allow for adiabatic
evolution and the system starts to populate the gapped mode: the
non-adiabatic dynamics  starts. Below we assume that the separation 
of the two regimes takes place at $q=2-\hat q$, and determine the
location of $\hat q$ as a function of the quench rate.
This can be done  as in Sec. \ref{sec1}.
Comparison of the  timescales (\ref{111}) and (\ref{222})
brings us to the (approximate) equation 
\begin{equation}
\label{gap_equation}
\frac{\hbar}{\Delta(\hat q)}=\left.\frac{\Delta}{|\dot\Delta|}\right|_{\hat
q}.
\end{equation}
 This equation is illustrated in Fig. \ref{gap_eq_new}. 
Its solution  in the slow transition limit, $\hat q\ll1$,  gives

\begin{equation}
\label{qhat}
\hat q\cong \frac{1}{4^{2/3}}\left(\frac{\tau_0}{\tau_Q}\right)^{2/3}, \ \ 
\tau_0=\frac{\hbar}{n|c_1|}.
\end{equation}
This, of course, corresponds to (\ref{lambda_hat}) with $z\nu=1/2$ and 
$\Delta_0= 2n|c_1|$.
A simple estimation of the constant $\tau_0$ by the peak density 
in the Berkeley experiment, $2.8\times10^{14}cm^{-3}$, gives
$\tau_0\approx16 ms$. In this section and Sec. \ref{sec4} all
the results can be obtained without assuming any particular value
of $\tau_0$.

Qualitatively, prediction (\ref{qhat}) is in  agreement with what we expect:
the slower the quench is (the larger $\tau_Q$ is), the closer to the
critical point the condensate can be driven adiabatically.

\begin{figure}
\includegraphics[width=0.5\columnwidth,clip=true]{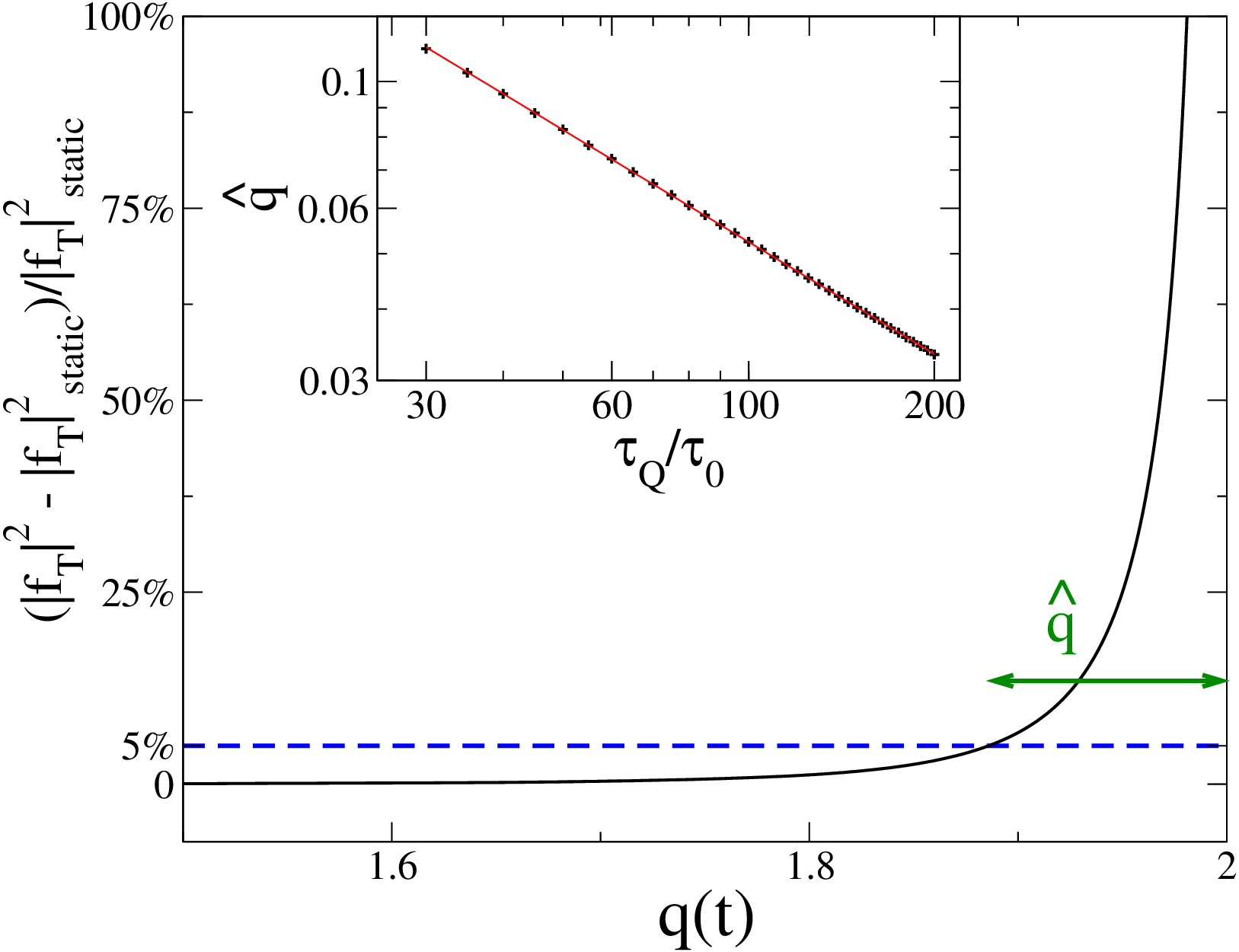}
\caption{(color) Departures of the transverse magnetization from the static 
         prediction during evolution in the broken-symmetry phase.
	 $|f_T|^2_{\rm static}$ refers to the squared modulus of 
	 (\ref{fT_}).
	 We arbitrarily define that when these departures exceed 
	 $5\%$, the system leaves the adiabatic regime and starts
	 the non-equilibrium dynamics (other reasonable thresholds $<10\%$
	 give the same scaling result). The inset shows the result
	 of power law fit pointing to  
	 $\hat q=1.05\times(\tau_0/\tau_Q)^{0.651\pm0.001}$, 
	 which compares well to theoretical prediction (\ref{qhat}). 
	 The fit is presented as a solid line, while numerics 
	 comes as pluses. Note the log-log scale on the inset.
	 The fit was done for $\tau_Q/\tau_0=30\cdots200$. The larger
	 $\tau_Q$'s are taken to the fit, the closer scaling exponent
	 approaches $2/3$. 
	 On the main plot we have a simulation for $\tau_Q/\tau_0=30$.
}
\label{depart}
\end{figure}

Quantitatively, solution (\ref{qhat}) turns out to be remarkably 
accurate in the wide range of quench times $\tau_Q$ as depicted in 
Fig. \ref{depart}.
In that plot we define the instant $\hat q$ to be
the distance  from a critical point when the
departure of system transverse magnetization from a static (ground state)
prediction starts exceeding some threshold, e.g., $5\%$
of the instantaneous GS value. From the fit there we get that 
$$
\hat q= 1.05\times\left(\frac{\tau_0}{\tau_Q}\right)^{0.651\pm0.001}
$$ 
in excellent agreement with (\ref{qhat}) with respect to the scaling exponent.
The prefactor, which obviously depends on the threshold used for
the numerical determination of $\hat q$ is of the same 
order as in (\ref{qhat}). Additional details are seen in Fig. \ref{depart}.
The error in the determination of the scaling exponent 
is one standard deviation coming from a linear fit on a log-log plot. 
All fitting errors in this paper are determined in this way.

\begin{figure}
\includegraphics[width=0.5\columnwidth,clip=true]{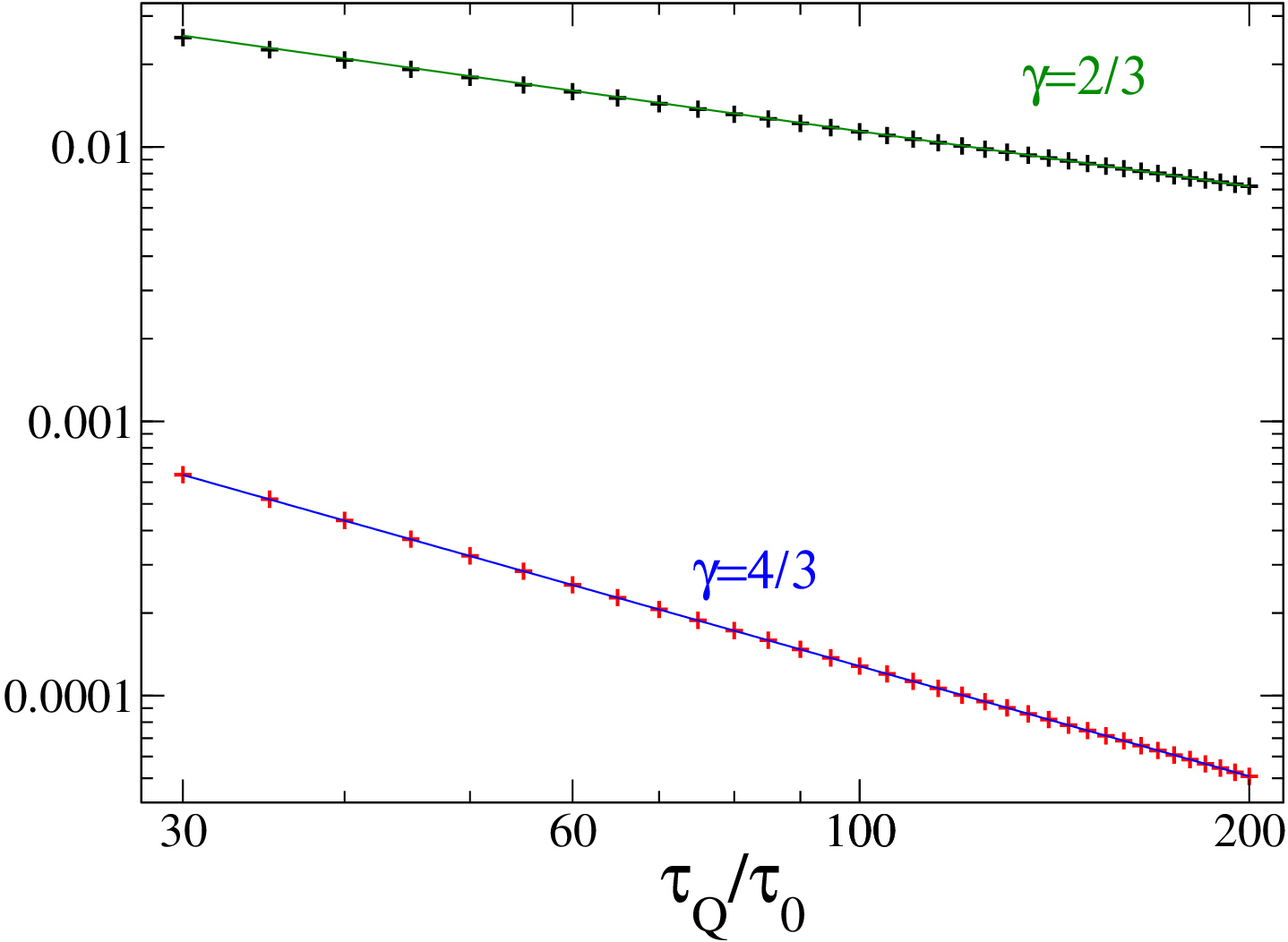}
\caption{(color) Scaling of the condensate properties at the {\it critical point}.
                 The upper data presents $|f_T|^2/n^2$, where $f_T$
		 is the transverse magnetization,
                 while the lower data is for $E/n^2|c_1|$, where
		 $E$ (energy density) is given by (\ref{E}). 
		 The pluses 
		 come from numerics, while the  solid lines 
		 present the curves $\sim(\tau_0/\tau_Q)^\gamma$
		 with $\gamma=2/3$ ($4/3$) for 
		 the upper (lower) data.
		 This confirms that at the critical point 
		 $|f_T|^2\sim\tau_Q^{-2/3}$, while $E\sim\tau_Q^{-4/3}$. 
}
\label{Emag}
\end{figure}

Let's look at the system magnetization at the critical point. 
To proceed further it is convenient to define the condensate
energy density as 
\begin{equation}
\label{E}
E=-\frac{|c_1|}{2}|f_T|^2+2n|c_1|q|\psi_1|^2,
\end{equation}
which is equal in the SMA to the contribution of the last two 
terms in Eq. (\ref{eeEE}) to the condensate  energy per volume (the first 
two terms in Eq. (\ref{eeEE}) 
are unimportant in the SMA as the gradient term equals zero
while the density-density interaction term $\sim c_0$ is 
a constant). 
In Fig.  \ref{Emag} we observe that 
$|f_T|^2\sim\tau_Q^{-2/3}$ and $E\sim\tau_Q^{-4/3}$. Putting these
two scalings into Eq. (\ref{E}) we see that 
$$
|\psi_1|^2= \frac{|f_T|^2}{8n}
$$
{\it at} the critical point in the leading, i.e., $\tau_Q^{-2/3}$,
order in the quench rate. 

\begin{figure}
\includegraphics[width=0.5\columnwidth,clip=true]{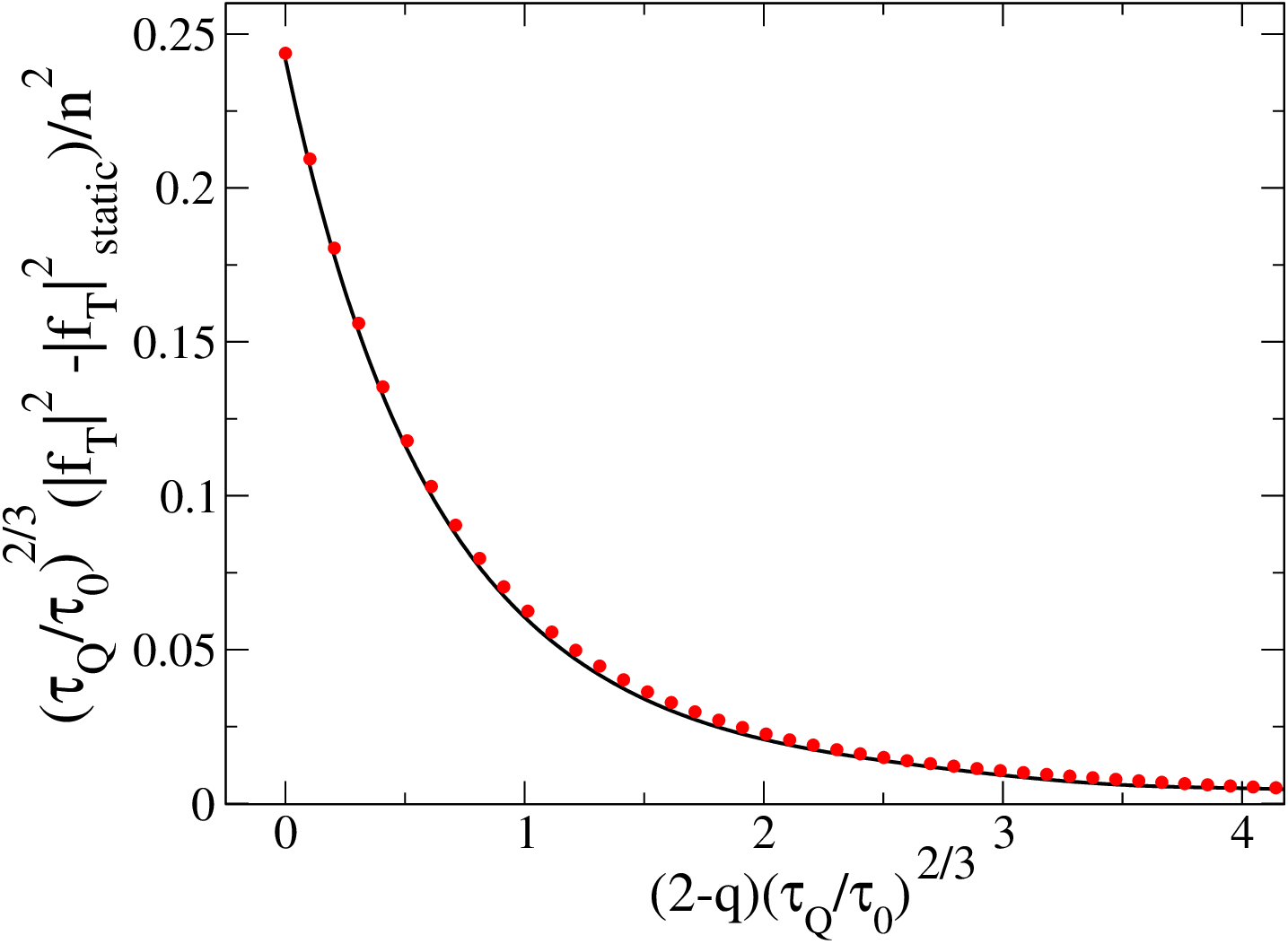}
\caption{(color) Illustration of the scaling of departures of the
         condensate transverse magnetization, $|f_T|^2$, 
	 from the static prediction, $|f_T|^2_{\rm static}$, given by 
	 the squared modulus of (\ref{fT_}).
	 The solid black line is for $\tau_Q/\tau_0=30$, while 
	 red dots show data for $\tau_Q/\tau_0=60$. 
	 The plot shows that the departures of the transverse magnetization 
	 from a GS prediction near a critical point scale as $\tau_Q^{-2/3}$  
	 after rescaling of the distance from the critical point, $2-q$, by 
	 a $\tau_Q^{2/3}$ factor.
}
\label{dfdq}
\end{figure}

It is now interesting to ask 
whether there is any relation between the condensate magnetization
at the point where the  system
goes out of equilibrium, $q=2-\hat q$, 
and the condensate  magnetization at the critical point, $q=2$. 
The condensate magnetization at $q=2-\hat q$ equals (in the slow transition 
limit where $\hat q\ll1$)
$$
|f_T|^2/n^2= 1- (2-\hat q)^2/4\approx\hat q\sim\tau_Q^{-2/3}
$$
so  $|f_T(q=2-\hat q)|^2$ scales with the quench rate in the same way 
as $|f_T(q=2)|^2$. This resembles  the adiabatic-impulse simplification
of dynamics \cite{bodzio,dorner}, where it is assumed that the evolution of 
a system undergoing a phase transition is either adiabatic or impulse,
and  the impulse stage implies no changes in the state of the system.
Here this simple picture gives a correct scaling of the
system magnetization 
at the critical point from the value of $\hat q$ -- the location where the 
system leaves the adiabatic regime. This is a good approximation  
despite the fact that the condensate  
does not enter an ideal ``impulse'' regime. Indeed, 
the  wave-function is not ``frozen'' and does change 
from $q=2-\hat q$ to $q=2$. Remarkably, 
however, these changes are such that 
 instead of getting 
 magnetization equal to $|f_T(q=2-\hat q)|^2$ 
 at the critical point  (ideal impulse dynamics) we get 
$const\cdot|f_T(q=2-\hat q)|^2$, where the constant 
is $\tau_Q$-{\it independent}. This 
character of  condensate evolution toward a critical point is 
{\it universal} with respect to quench 
time. It is depicted in Fig. \ref{dfdq}.

Now we would like to address how these critical scalings can be experimentally
observed. One obvious way is to watch the condensate dynamics
at different times as was done in the Berkeley experiment \cite{nature_kurn}. 
Another approach involving interesting physics 
would be to stop a change of $q$ and watch
subsequent magnetization dynamics. This approach will be discussed in the next
section.

\section{Dynamics of magnetization after the quench}
\label{sec4}

Suppose that one stops driving the system at some $q\ge2$ and
lets it evolve freely, which leads to  magnetization oscillations. 
We would like to investigate here whether 
the information about the system excitation at the critical point 
can be extracted from the amplitude and/or period of these
oscillations. In other words, we want to find out
what are the signatures of the nonequilibrium dynamics in the broken-symmetry
phase that may be observable in the polar phase. This problem 
is of both experimental and theoretical interest.

\subsection{Quench stops at the critical point}

\begin{figure}
\includegraphics[width=0.5\columnwidth,clip=true]{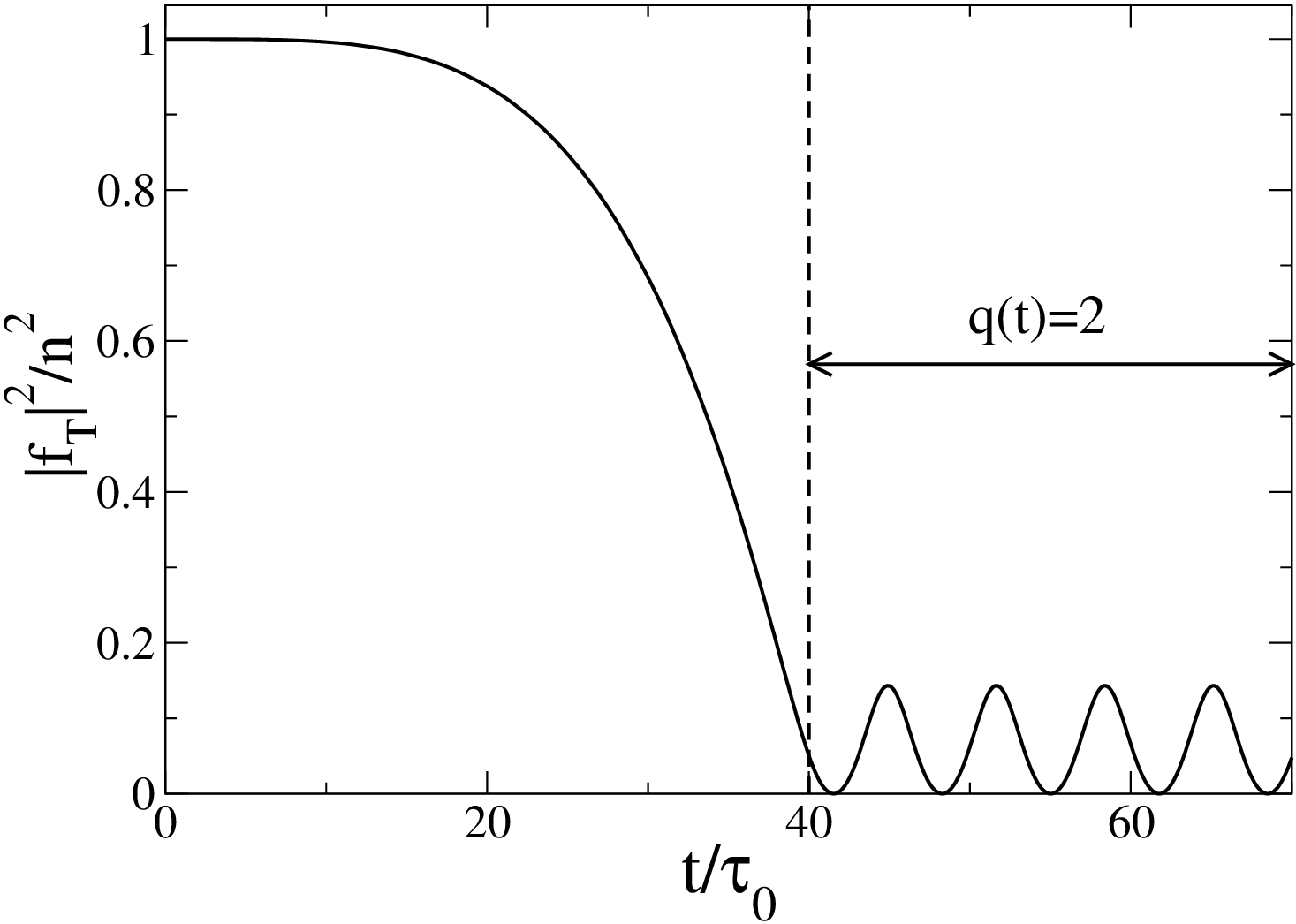}
\caption{ Time evolution of the magnetization during the 
          quench that stops at the critical point.
          This plot is for $\tau_Q/\tau_0=10$. 
	  The critical point is reached at $t/\tau_0=4\tau_Q/\tau_0=40$, so
	  $q(t\le40\tau_0)$ is given by (\ref{q_od_t}) 
	  while $q(t>40\tau_0)=2$. The system excitation due to 
	  approaching the critical point is 
	  visible from the nonzero magnetization at 
	  $t/\tau_0\ge40$: if the evolution would be adiabatic, the 
	  condensate magnetization at the critical point would be zero
	  and no magnetization oscillations would be present on the plot.
}
\label{polar_stop}
\end{figure}

Suppose the condensate is driven from $q=0$ to the critical point with 
(\ref{q_od_t}). The critical point is reached at $t=4\tau_Q$ and then 
$q(t\ge4\tau_Q)=2$, i.e.,  free (without any driving) 
evolution takes place. The evolution of magnetization 
for such a problem is presented
in Fig. \ref{polar_stop}, where periodic 
oscillations are easily seen. Our aim is to describe them and
show that both their amplitude and their period contain information 
on how the condensate was excited at the critical point.

In this case one can derive the following equation for the evolution of
the magnetization:
\begin{equation}
\label{qwerty}
\ddot{f}_T(t)=-\left(\frac{4|c_1|}{\hbar^2}E+ 2\frac{|c_1|^2}{\hbar^2}|f_T(t)|^2\right)f_T(t),
\end{equation}
where $E$ is the system energy density (\ref{E}) at the critical point (in fact, also 
at any other time since lack of parameter changes results in 
energy conservation). This equation can be obtained straightforwardly from 
Eqs. (\ref{fT}) and (\ref{3gp}). It is an interesting equation: 
in the limit of slow transition, when the amplitude of oscillations 
becomes very small, the nonlinear term dominates the 
physics rather than being negligible compared to
the $\sim E$ term promoting 
harmonic oscillations. Indeed, from Sec. \ref{sec3} we see that 
$|c_1|E/|c_1|^2|f_T|^2\sim(\tau_0/\tau_Q)^{2/3}\ll1$ for slow 
enough quench.

The exact solution  reads 
\begin{equation}
\label{cosineJ}
f_T(t)= A\, {\rm cn}\left(\frac{|Ac_1|}{\hbar\kappa}t+\theta,\kappa\right),
\end{equation}
where $\rm cn$ is the Jacobi cosine \cite{whittaker}, 
$$
\kappa= \sqrt{\frac{\alpha}{1+\alpha}}\frac{1}{\sqrt{2}}, \ \ \alpha=
\frac{|c_1|}{2E}|A|^2,
$$
and $A$ and $\theta$ are given by initial conditions. They both might
be found from a fit to experimental data. In particular, $|A|^2$ is
given by the amplitude of free $|f_T|^2$ oscillations, because 
the Jacobi cosine (\ref{cosineJ}) oscillates periodically between 
$\pm1$.

\begin{figure}
\includegraphics[width=0.5\columnwidth,clip=true]{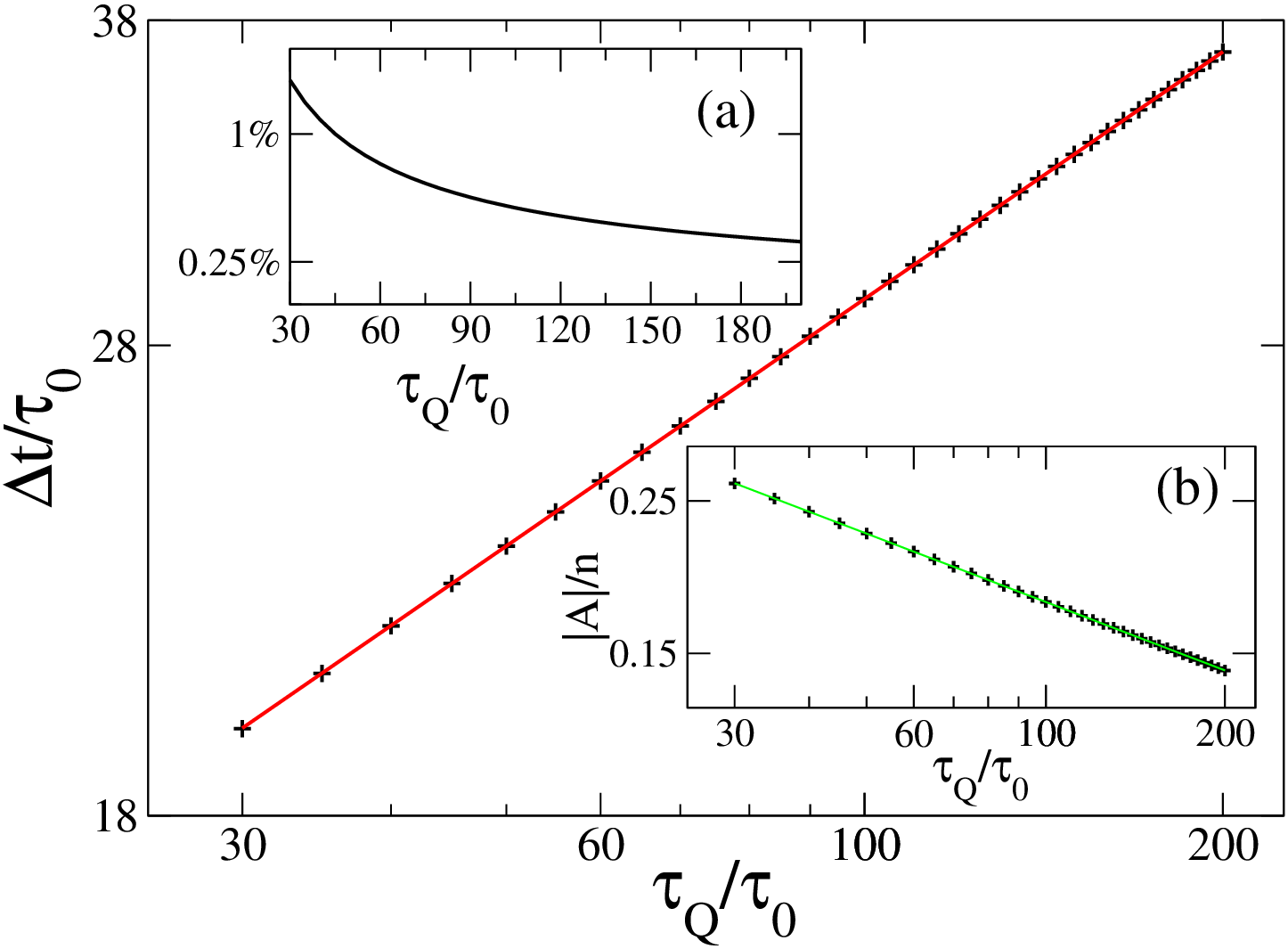}
\caption{(color) Here we show the properties of the system when the change of $q$
         stops at the critical point and then there is a free evolution
	 going on. Note that there is a logarithmic scale on the
	 vertical and horizontal axes of the main plot and the inset (b).
	 Main plot: the periodicity interval 
	 $\Delta t$ of $f_{x,y}(t)$ during free oscillations. The fit (red line)
	 to numerics (pluses) gives  $\Delta t\sim \tau_Q^{0.3350\pm0.0001}$. 
	 Inset (a): the  difference 
	 between (\ref{DT}) and (\ref{full_DT}) divided by (\ref{full_DT}),
	 i.e., relative departures of the approximate formula for oscillation
	 period from the exact result. The approximate
	 formula (\ref{DT}) overestimates the exact result by about one percent 
	 in a wide range of $\tau_Q/\tau_0$'s.
	 Inset (b):  dependence of $|A|$ from (\ref{cosineJ}) on
	  quench time $\tau_Q$. The fit (green line) to numerics
	 (pluses) gives $|A|\sim\tau_Q^{-0.3304\pm0.0001}$ in agreement with
	 adiabatic - impulse simplification of the condensate dynamics.
}
\label{dt}
\end{figure}

The Jacobi cosine was formerly used in the spin-1 condensate context in 
\cite{ferro_exp_jacobi,cn}. It is periodic --
${\rm cn}(x,\kappa)={\rm cn}(x+4K(\kappa),\kappa)$ --
where $K(\kappa)$ is the complete elliptic integral of the first kind. In the time domain 
this periodicity translates into  
\begin{equation}
\label{full_DT}
\Delta t= 4\kappa\left|\frac{\hbar}{Ac_1}\right|K(\kappa).
\end{equation}
To simplify this expression we note that from the fact that 
the Jacobi cosine in (\ref{cosineJ}) is bounded by $\pm1$,
$f_T$ and $A$ are of the same order and so we expect 
$|A|\sim\tau_Q^{-1/3}$. Moreover, as $E\sim\tau_Q^{-4/3}$
we get that in the limit of large $\tau_Q$ (slow transitions) 
$\alpha\sim(\tau_Q/\tau_0)^{2/3}\gg1$
and so $\kappa\to1/\sqrt{2}$. This last result allows us to write
\begin{equation}
\label{DT}
 \Delta t= \frac{\Gamma(1/4)^2}{\sqrt{2\pi}}\left|\frac{\hbar}{Ac_1}\right|.
\end{equation}
This finding is quite interesting: the periodicity is amplitude-dependent
unlike in the harmonic oscillator case. This simple result predicting the
inverse proportionality of the period and the amplitude of oscillations is
accurate to about $1\%$ for experimentally relevant quenches as depicted in 
inset (a) of Fig. \ref{dt}.

The numerical results on the oscillation period $\Delta t$ when the quench
stops at $q=2$ are presented in Fig. \ref{dt}. From the fit we see that 
 $\Delta t$ scales as $\tau_Q^{1/3}$ in accordance with 
expression (\ref{DT}) supplemented with the above observation that 
$|A|\sim\tau_Q^{-1/3}$ (proven numerically in the inset (b) of Fig. \ref{dt}).

\subsection{Quench stops after passing  the critical point}
\begin{figure}
\includegraphics[width=0.5\columnwidth,clip=true]{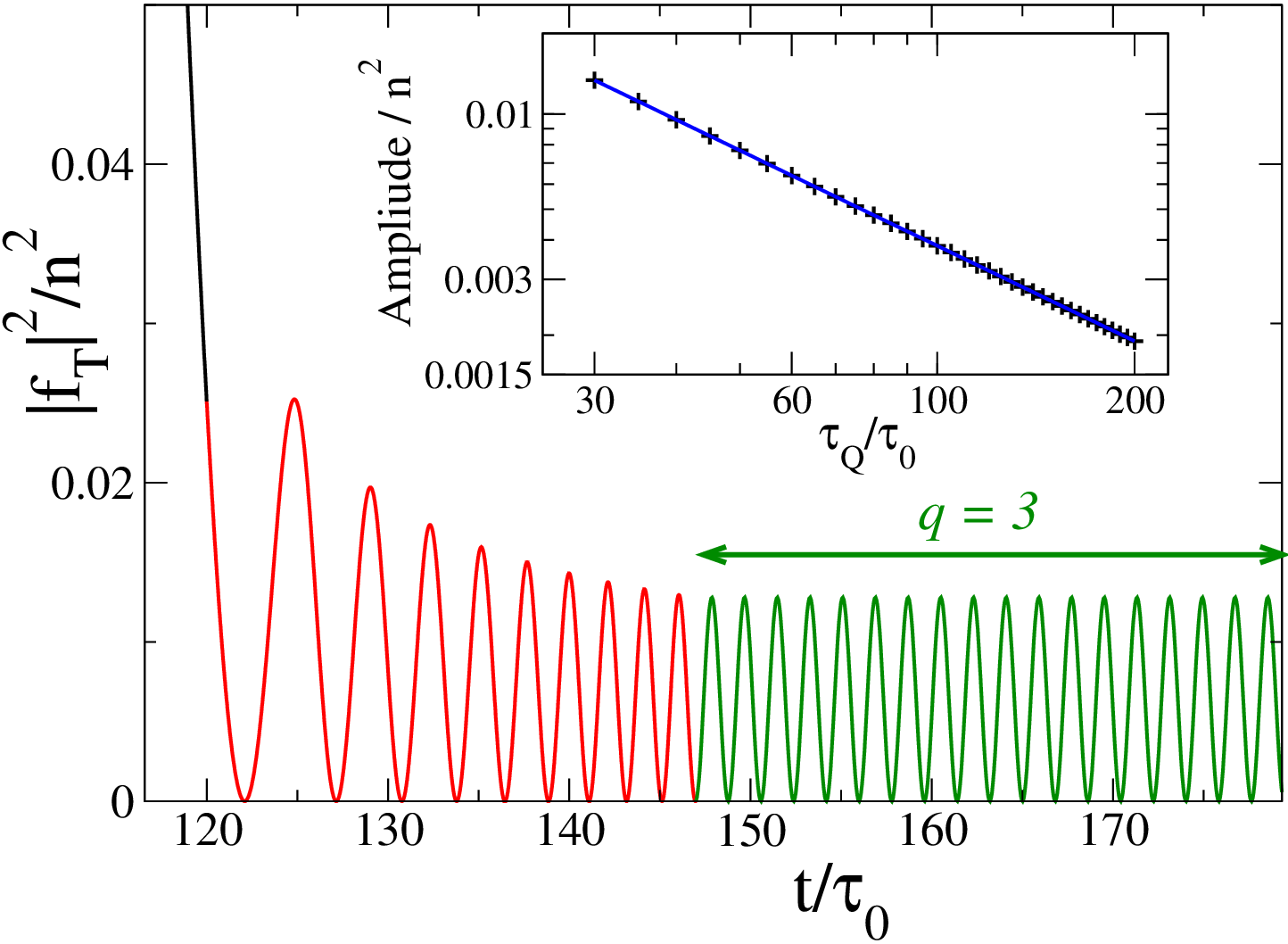}
\caption{(color) Dynamics of the system driven by the change of $q(t)$ 
         given by (\ref{q_od_t}) until the point $q=3$,
	 and then undergoing a free 
	 evolution with $q$ time independent. 
	 The main plot 	 shows data for $\tau_Q/\tau_0=30$.
	 The free evolution  starts at 
	 $t/\tau_0=\sqrt{24}\tau_Q/\tau_0\approx147$.
	 Different stages of evolution
	 are encoded in colors: black (evolution in broken-symmetry phase),
	 red (dynamics in polar phase induced by change of $q$), green
	 (free dynamics in the polar phase, $\dot q=0$).
	 Inset: the scaling of the amplitude of $|f_T(t)|^2$
	 oscillations during free evolution at $q=3$ (depicted in green). 
	 Pluses come from numerics, while the line is the fit corresponding to 
	 $\sim\tau_Q^{-1.0013\pm0.0001}$ scaling.
}
\label{dampl}
\end{figure}

According to our numerical simulations there are three stages of such evolution 
depicted in Fig. \ref{dampl}: (i)
the system is driven 
to the critical point through the entire broken symmetry phase
and its magnetization at the critical point scales as
$|f_T|^2\sim\tau_Q^{-2/3}$ (only barely noticeable 
magnetization oscillations are present -- see Ref. \cite{liniowe}); 
(ii) the condensate is driven by a change of $q$,
Eq. (\ref{q_od_t}), in the polar phase and we observe there damped magnetization
oscillations; (iii) the free evolution after we stop ramping up 
$q(t)$ takes place and periodic magnetization oscillations appear.
These oscillations are described by 
\begin{equation}
\label{xxx}
\ddot{f}_T=-\left[\tau_0^{-2}q(q-2) 
	   + 4\frac{|c_1|}{\hbar^2}E+ 2\frac{|c_1|^2}{\hbar^2}|f_T(t)|^2\right]f_T,
\end{equation}
where again $\dot E=0$ and Eqs. (\ref{fT}) and (\ref{3gp}) have been
employed in derivation of (\ref{xxx}). 
The exact solution in terms of the Jacobi cosine function exists and has
 form (\ref{cosineJ}) with 
$$
\alpha=  \frac{|Ac_1|^2}{q(q-2)\hbar^2/2\tau_0^2+2E|c_1| }.
$$
Now, in the limit of fixed $q>2$ and large $\tau_Q$ (slow transition)
we have a different behavior  of the Jacobi cosine than above. Namely,
now $\alpha\sim (\tau_0/\tau_Q)^{2/3}\ll1$ which results in $\kappa\to0$
and  ${\rm cn(\bullet,\kappa)}\to\cos(\bullet)$. 
In this limit the last two terms in (\ref{xxx}) 
can be neglected and the Jacobi cosine turns into a normal cosine function: 
harmonic oscillations show up with a repetition period 
\begin{equation}
\label{DT_osc}
\Delta t= \frac{2\pi\tau_0}{\sqrt{q(q-2)}}.
\end{equation}
Naturally, the transition between Jacobi cosine oscillations and
typical harmonic dynamics is gradual and can be traced quantitatively with 
the exact solution (\ref{cosineJ}). We can compare (\ref{DT_osc}) 
to the numerics. For the evolutions where the increase of $q$ stops 
at $q=3$, as is the case in Fig. \ref{dampl}, we get from (\ref{DT_osc})
that $\Delta t=2\times1.814\tau_0$, while from numerics we  obtain
$2\times1.804\tau_0$ for $\tau_Q/\tau_0=30$ and $2\times1.812\tau_0$ for
$\tau_Q/\tau_0=200$ (notice that 
$|f_T|^2$ oscillates with $1/2$ of the period (\ref{DT_osc}) and so 
our numerical results extracted from $|f_T|^2$ oscillations
were multiplied by a factor of two). 
As we see, the larger $\tau_Q$, the better the agreement because the nonlinear
corrections become  smaller. This is in fact  bad news: as (\ref{DT_osc}) 
is $\tau_Q$-independent, we can no longer 
use repetition period  $\Delta t$ to investigate the dynamics of 
the quantum phase transition. 

Fortunately, however, behavior of the {\it amplitude} of free 
magnetizations  oscillations  provides an easily visible
signature of the non-equilibrium dynamics. 
As shown in the inset of Fig. \ref{dampl},
the amplitude of $|f_T|^2$ scales as $\tau_Q^{-1}$. 
This scaling results from two observations. First,
the driving in the polar phase starting from $q=2$ and
proceeding to $q\gtrsim3$ damps the amplitude of
$|f_T|^2$ by a factor of $\tau_Q^{-1/3}$. 
We have verified this numerically by simulating different evolutions that 
begin from a fixed, $\tau_Q$ independent, state. Second, the scaling of $|f_T|^2$
at the phase boundary is $\tau_Q^{-2/3}$. Combining these two 
results one easily justifies $\tau_Q^{-1}$ scaling from the numerics.

\begin{figure}
\includegraphics[width=0.5\columnwidth,clip=true]{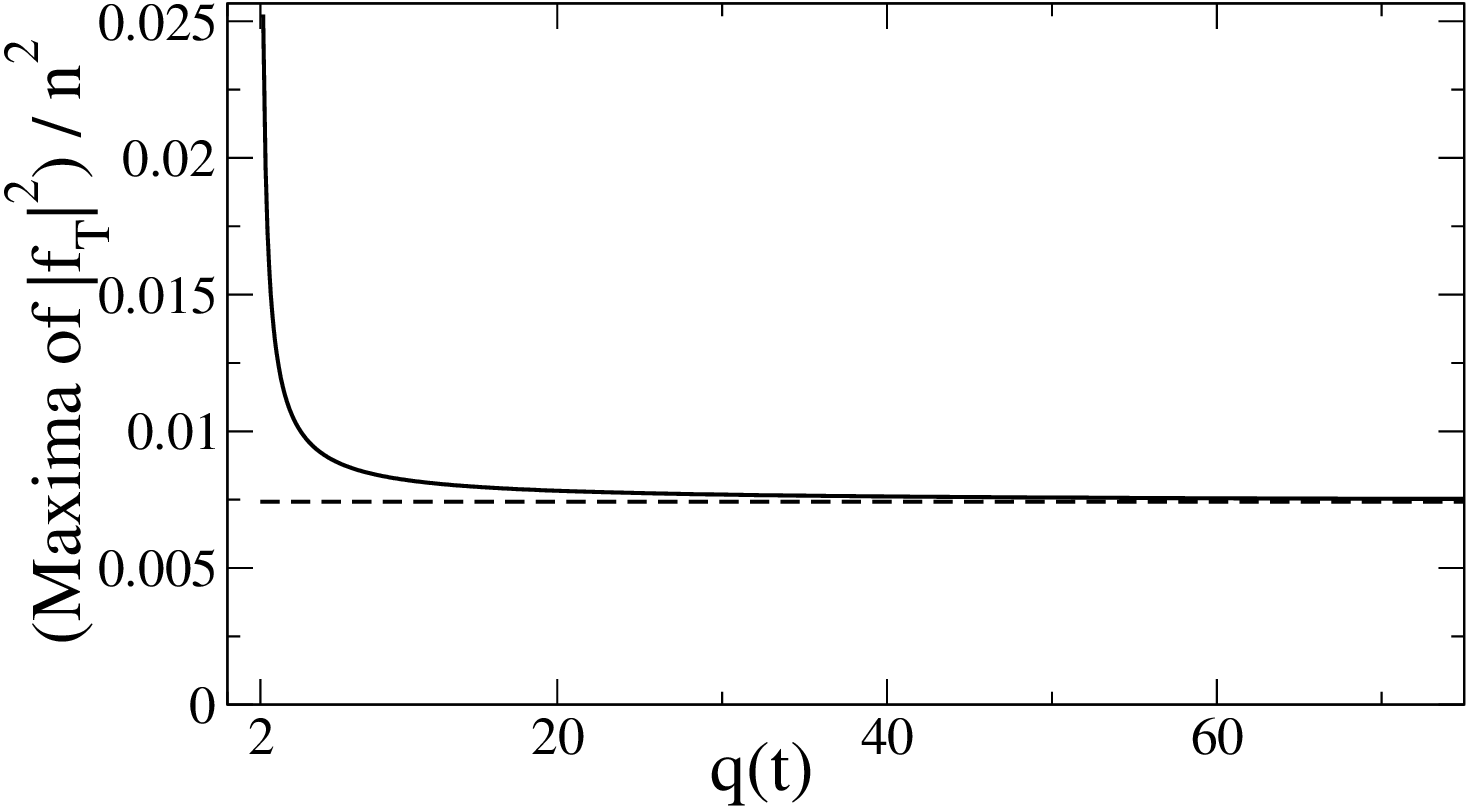}
\caption{ 
          Maxima of the system magnetization, $|f_T|^2$,
          during an uninterrupted driving through
	  the polar phase with $q(t)$ given by  (\ref{q_od_t}).
	  The solid line comes from numerics and provides the upper envelope for
	   oscillations of $|f_T(t)|^2$.
          The data is for one run with  $\tau_Q/\tau_0=30$. 
	  The dashed line is the asymptotic (large $q$)
	  value for the solid line. It is obtained from a 
	  linear fit $({\rm Maxima \ of \ } |f_T|^2)/n^2 = a + b\frac{1}{q}$ for 
	  $q\sim100$ data. The fit gives $a\approx0.0074$ 
	  plotted as a dashed line. Additionally, we have verified numerically, 
	  that for different $\tau_Q$'s the following scaling holds:
	  $a\sim\tau_Q^{-1}$, which can be explained in the same way as 
	  the fitting result from the inset of Fig. \ref{dampl}.
}
\label{omega}
\end{figure}

Finally, we comment on what happens when the system is continuously
driven through  the  polar phase. Combining Eqs. (\ref{fT}) and (\ref{3gp})
one easily arrives at  
\begin{equation}
\label{full_diff}
\ddot{f}_T = -\tau_0^{-2} q(t)(q(t)-2)f_T
               -8\frac{n|c_1|^2}{\hbar^2}q(t)|\psi_1(t)|^2f_T +\frac{\dot q(t)}{q(t)}\dot f_T,
\end{equation}
that has to be solved with initial conditions given at instant $t_i$ by 
$f_T(t_i)$ and 
$\dot{f}_T(t_i)=iq\sqrt{2}\tau_0^{-1}(\psi_1^*\psi_0-\psi_{-1}\psi_0^*)|_{t_i}$.
Two remarks are in order. First, the same initial conditions apply to 
(\ref{qwerty}) and (\ref{xxx}). Second,
equation (\ref{full_diff}) is no longer self-consistent: an additional 
equation for $\psi_1(t)$ dynamics has to be solved simultaneously 
unless the term $\sim|\psi_1(t)|^2$ is negligible.

As we have observed in Fig. \ref{dampl}
the magnetization oscillations are damped when the system is driven 
in the polar phase. It turns out, however, that when we continue 
driving the condensate, the amplitude of magnetization oscillations reaches some nonzero 
asymptotic value, scaling as $\tau_Q^{-1}$, instead of decreasing to zero. 
This is  
supported by  Fig. \ref{omega} and the following analysis.
First, we assume that 
$q(t)\gg2$ such that it is safe to approximate $q(t)(q(t)-2)$
by $q(t)^2$. Second, we neglect the second term in (\ref{full_diff}) 
as it is small compared to the first one  in the 
slow transition limit. After that we end up with the following equation
that can be solved exactly
\begin{equation}
\label{small_diff}
\ddot{f}_T= -\tau_0^{-2}q(t)^2f_T + \frac{\dot q(t)}{q(t)}\dot f_T.
\end{equation}
This is precisely a driven harmonic oscillator with {\it anti-friction} since 
$\dot q(t)/q(t)$ is positive in our case. In a classical mechanical
picture we may imagine that (\ref{small_diff}) describes a particle in a 
harmonic potential under the influence of a force that 
acts along its velocity. Such a force fights the tendency to squeeze particle
motion due to increase of the frequency of the harmonic oscillator. Interestingly,
it turns out that the particle perfectly maintains the amplitude of its
motion because the exact solution of (\ref{small_diff}), constrained by 
$\dot\phi=0$ from (\ref{3gp}), is 
$$
f_T= C\cos\left(\frac{1}{\tau_0}\int dt q(t)+\theta\right), \ \ \dot C=0.
$$
Additionally, we stress the fact that {\it this solution is valid for
any smooth $q(t)$ dependence} and not just (\ref{q_od_t}). 
In particular, in the opposite limit when $q(t)$ decreases in time
the ``anti-friction'' term works as a special friction term that slows the
particle down so that it oscillates with constant amplitude instead of
spreading out.

This phenomenon can be qualitatively explained as follows. As it happens at $q(t)\gg2$
we can safely assume that the magnon modes of the system 
do not become excited any more  during driving. 
This is so because they have energy gap that becomes large far away from the critical
point.
These modes are responsible for the scattering of atoms 
between  $m=0$  and $m=\pm1$ condensate components \cite{ueda_spectrum}. 
In the absence
of that process the populations of $m=0,\pm1$ sublevels should be  
constant, which implies that 
$$
|f_T|^2 = 8 |\psi_1|^2|\psi_0|^2 \cos^2\left(\chi_0-\frac{\chi_1+\chi_{-1}}{2}\right)
$$
will undergo {\it fixed} amplitude oscillations induced by rotation 
of the condensate phases.

\section{Dynamics in a slightly inhomogeneous system}
\label{sec5}

In this section we would like to compare the results obtained from a 
Single Mode Approximation enforcing the translational symmetry onto 
wave-function of the system  to the more experimentally relevant inhomogeneous
(``disordered'') problem. To this aim we  assume that the atom cloud
is placed in a box-like trap and impose density and phase fluctuations
onto it. These may come from experimental imperfections and 
quantum  fluctuations.

The results presented here come from numerics done in a one-dimensional
configuration and so we present an explicitly 1D description
below unless stated otherwise. 
We consider untrapped system: atoms in the box as in 
the experiment \cite{raizen} done with spinless bosons.
We estimate the parameters for
these 1D simulations as in \cite{ferro_bodzio} 
taking into account the experimental setup  of \cite{nature_kurn}.
Below $L\approx400\mu m$ is the size of the box keeping $N=2\times10^6$ 
atoms confined along the $z$ direction.

\begin{figure}
\includegraphics[width=0.5\columnwidth,clip=true]{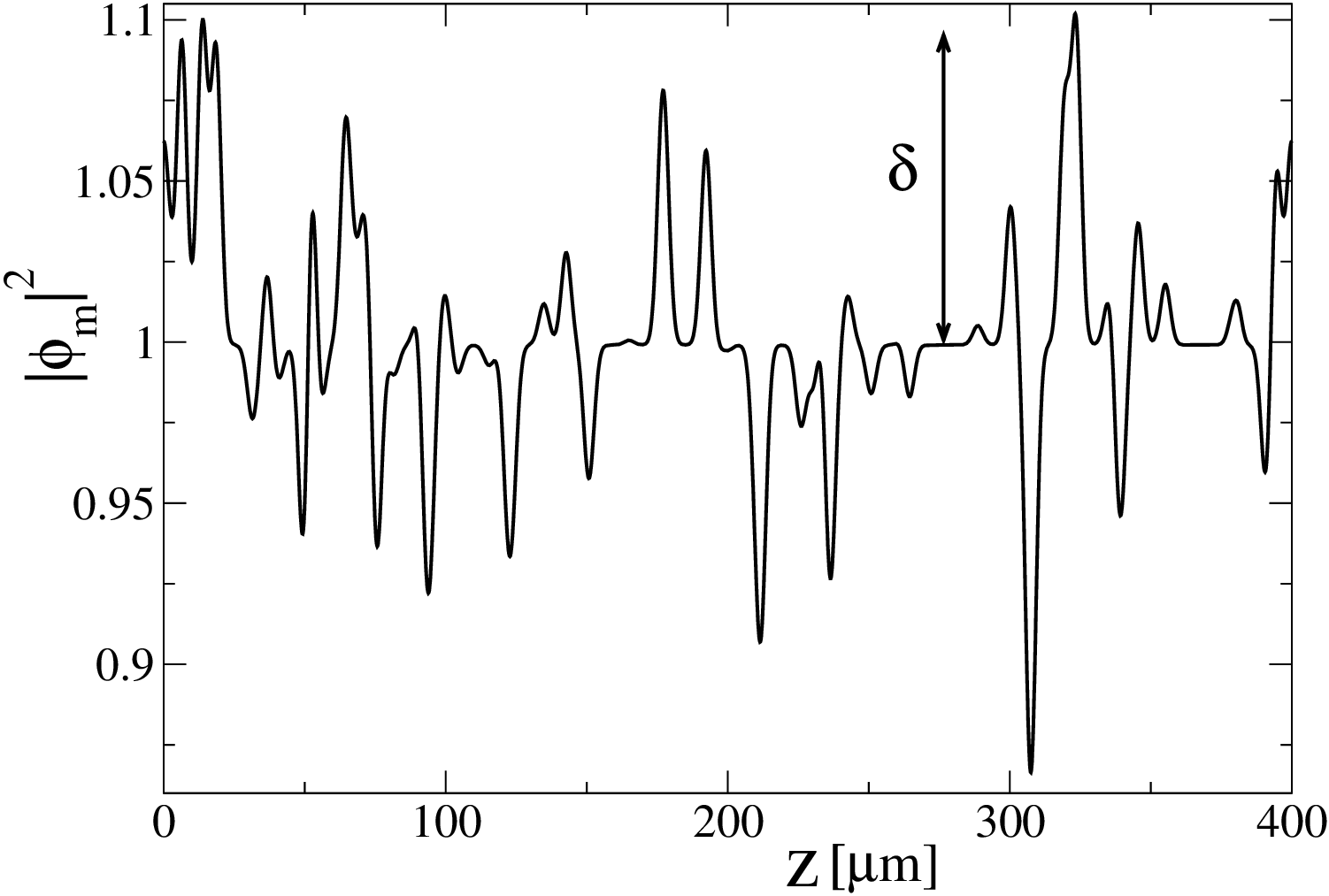}
\caption{Plot of $|\phi_m(z)|^2$ for a typical realization of disorder
(\ref{disorder}). The parameter $\delta=10\%$.
}
\label{disorder1}
\end{figure}

We introduce disorder by modifying the initial wave function at
$q(t=0)=0$ in the following way:
$$
\Psi\approx\sqrt{n}
\left(
\begin{array}{c}
\frac{1}{2}\phi_1(z) e^{i\chi_1} \\ 
\frac{1}{\sqrt{2}}\phi_0(z) e^{i(\chi_1+\chi_{-1})/2} \\ \frac{1}{2}\phi_{-1}(z) e^{i\chi_{-1}}
\end{array}
\right), \ \int_0^L dz |\Psi|^2= N, 
$$
where the mode amplitude is proportional to:
\begin{equation}
\label{disorder}
\phi_m=\sqrt{1 + \delta \sum_{n=1}^{\kappa} e^{i\eta_n^{(m)}} r_n^{(m)} 
						           \exp\left(-\frac{(z-z_n^{(m)})^2}{\sigma^2}\right) }.
\end{equation}
There $\delta$ gives the strength of the disorder,
$\eta^{(m)}_n,\chi_{\pm1}\in[0,2\pi)$ are random phases, $r_n^{(m)}\in[0,1]$ are random
amplitudes, and $z_n^{(m)}\in[0,L]$ are random positions of gaussian
disturbances. All these parameters are generated with uniform probability
density. We have chosen $\sigma$ to be equal to the spin healing length $\xi_s$
and assumed that the average spacing between Gaussian perturbations is
$2\xi_s$. The spin healing length in the experiment \cite{nature_kurn} is $2.4\mu
m$, so that $\kappa$, the number of Gaussians in (\ref{disorder}), 
is about $80$ ($L/2\xi_s$). 
The meaning of $\delta$ is that 
it provides a characteristic relative-to-background size of density 
variations. 
A typical disordered pattern in (\ref{disorder}) is
depicted in Fig. \ref{disorder1}.
Now our goal is to answer what is the range of the $\delta$ parameter
that allows for observation of qualitatively the same physics as in the
homogeneous system described in Secs. \ref{sec3} and \ref{sec4}.

To look at the dynamics in the broken-symmetry phase, 
we have done calculations for different $\delta\in(0,0.2)$. 
For each $\delta$ we randomly generate many-times the
pattern (\ref{disorder}) and evolve the system to 
the critical point  at different quench times $\tau_Q$. The patterns 
are independently generated for every evolution, which shall
resemble the experimental situation where the structure of ``imperfections''
fluctuates from run to run. Then for each pair $(\delta,\tau_Q)$  we 
average the mean system magnetization over $N_r=30$ runs, 
$$
\langle|f_T|^2\rangle=\frac{1}{N_r}\sum_{i=1}^{N_r}\frac{1}{L}\int_0^Ldz|f_T^{(i)}(z,t)|^2,
$$
where $f_T^{(i)}(z,t)$ is the condensate transverse magnetization in the $i$-th
run.

We look at two 
quantities: the distance $\hat q(\tau_Q)$ from the critical point where 
the system starts the  non-adiabatic evolution due to 
approaching the critical point, 
and the scaling of the condensate magnetization at the critical point.
Both are defined in the same way as in Sec. \ref{sec3} 
except that now we use $\langle|f_T|^2\rangle$ instead of $|f_T|^2$.
Our results suggest 
that the scaling exponents in the disordered system can
match the SMA predictions within a few percent accuracy for $\delta\lesssim2\%$.
Quantitatively, as we see in Table \ref{table}, 
the scaling exponent $\lambda$ given by  
$\hat q\sim\tau_Q^{-\lambda}$ stays reasonably close to 
the $2/3$ value for the homogeneous system 
for $\delta$ as large as $3\%$. For larger $\delta$, e.g., $\delta=5\%$ 
the numerical data clearly departs from the power law. 
The scaling exponent $\sigma$, defined at the critical point as  
$\langle|f_T|^2\rangle\sim\tau_Q^{-\sigma}$, deviates noticeably  
from the SMA prediction already for $\delta=2\%$ (Table \ref{table}). 
For larger $\delta$, e.g., $\delta=3\%$ numerics does not follow the power law anymore.

\begin{table}
\begin{tabular}{|c|c|c|}
\hline
          & $\lambda$  & $\sigma$    \\
\hline
$\delta=0.5\%$ & $0.65\pm0.01$   &   $0.650\pm0.001$ \\
\hline
$\delta=1\%$   & $0.67\pm0.01$   &   $0.647\pm0.001$ \\
\hline
$\delta= 2\%$  & $0.65\pm0.01$   &   $0.60\pm0.02$   \\
\hline
$\delta=3\%$   & $0.64\pm0.02$   &    		\\
\hline
\end{tabular}
\caption{The scaling exponents $\lambda$, $\hat q \sim \tau_Q^{-\lambda}$, and  $\sigma$,
$\langle|f_T|^2\rangle$ at the critical point $\sim\tau_Q^{-\sigma}$.
The values and errors (one standard
deviation) come from a linear fit on a log-log plot to data in the range of 
$\tau_Q=74ms\cdots1.1s$.
The Single Mode Approximation ($\delta=0$) points to
$\lambda,\sigma\approx2/3$. 
The definition of $\hat q$ is the same as in Fig. \ref{depart} except
$\langle|f_T|^2\rangle$ is used instead of $|f_T|^2$.
}
\label{table}
\end{table}

There are at least two reasons for discrepancies between 
disordered and homogeneous results. 
First and most importantly, 
the presence of disorder perturbs magnetization
affecting determination of both exponents from Table \ref{table}. 
Even when these departures from a 
GS are relatively small compared to the GS magnetization when the
system starts time evolution at $q=0$,  they can be significant
when compared to the condensate  magnetization close to or at the critical point
where scaling exponents are determined.
Second, our prediction of the scaling exponents 
is based on translationally invariant theory, i.e., momentum $k=0$
problem, while the introduction of disorder leads to population
of $k\neq0$ modes. A more involved description with the gap $\Delta$
(\ref{ddd})
being a function of $k$
should  be used when the population of $k\neq0$ modes becomes significant. 

\begin{figure}
\includegraphics[width=0.5\columnwidth,clip=true]{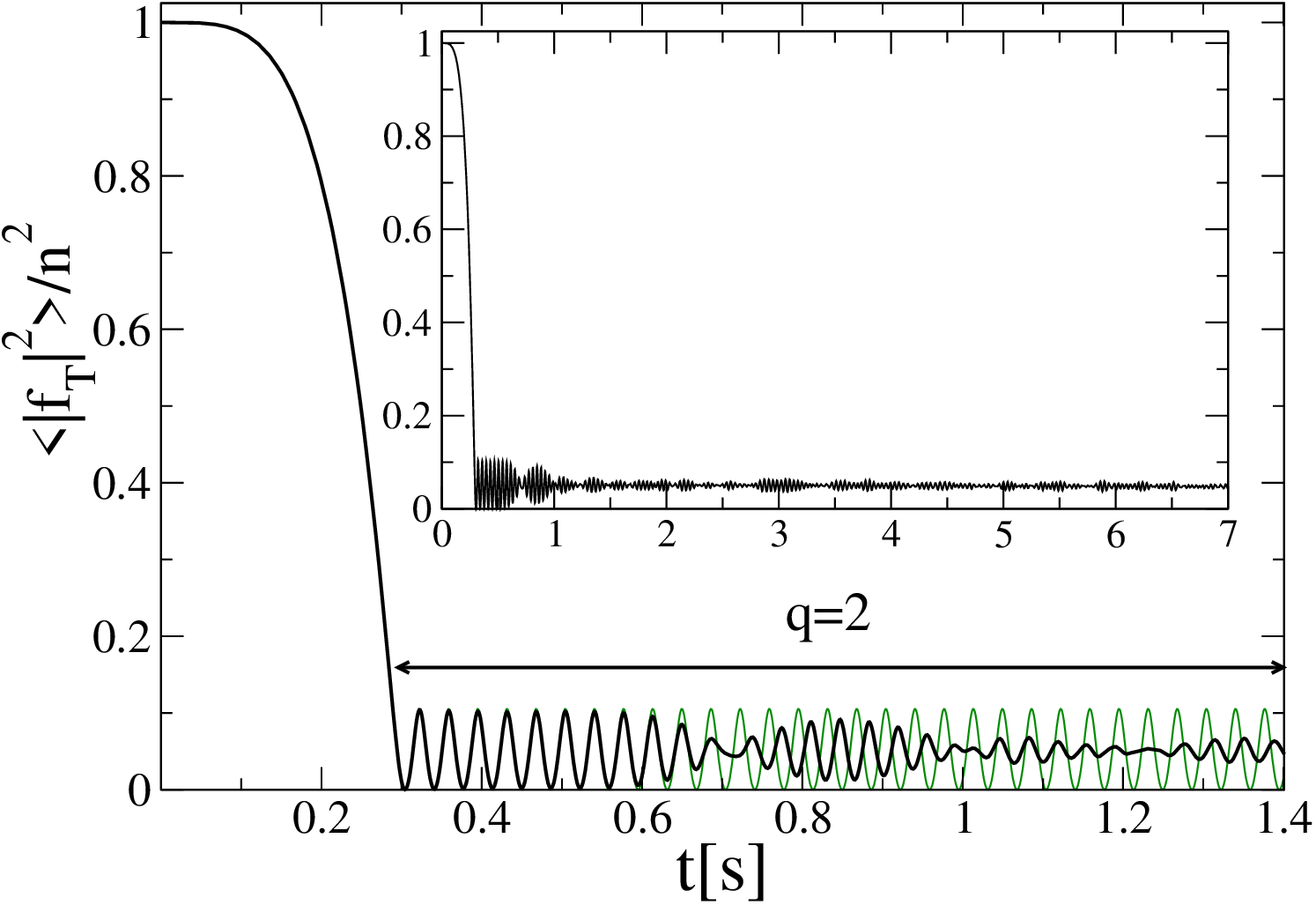}
\caption{(color) Dynamics of the 
         transverse magnetization in the inhomogeneous system during
         a quench with $\tau_Q=74ms$. The change of
         $q(t)$  stops at the critical point: $q(t\ge4\tau_Q\approx0.3s)=2$.
	 The thick black line is for $\delta=1\%$. 
	 The thin green
	 line presents $\delta=0$ result, i.e., the Single Mode Approximation 
	 outcome. The inset shows long time behavior of the 
	 black thick curve from the main plot.
	 The initial state for that evolution is the same as in Fig.
	 \ref{stop3_disorder}.
} 
\label{stop2_disorder}
\end{figure}

\begin{figure}[t]
\includegraphics[width=0.5\columnwidth,clip=true]{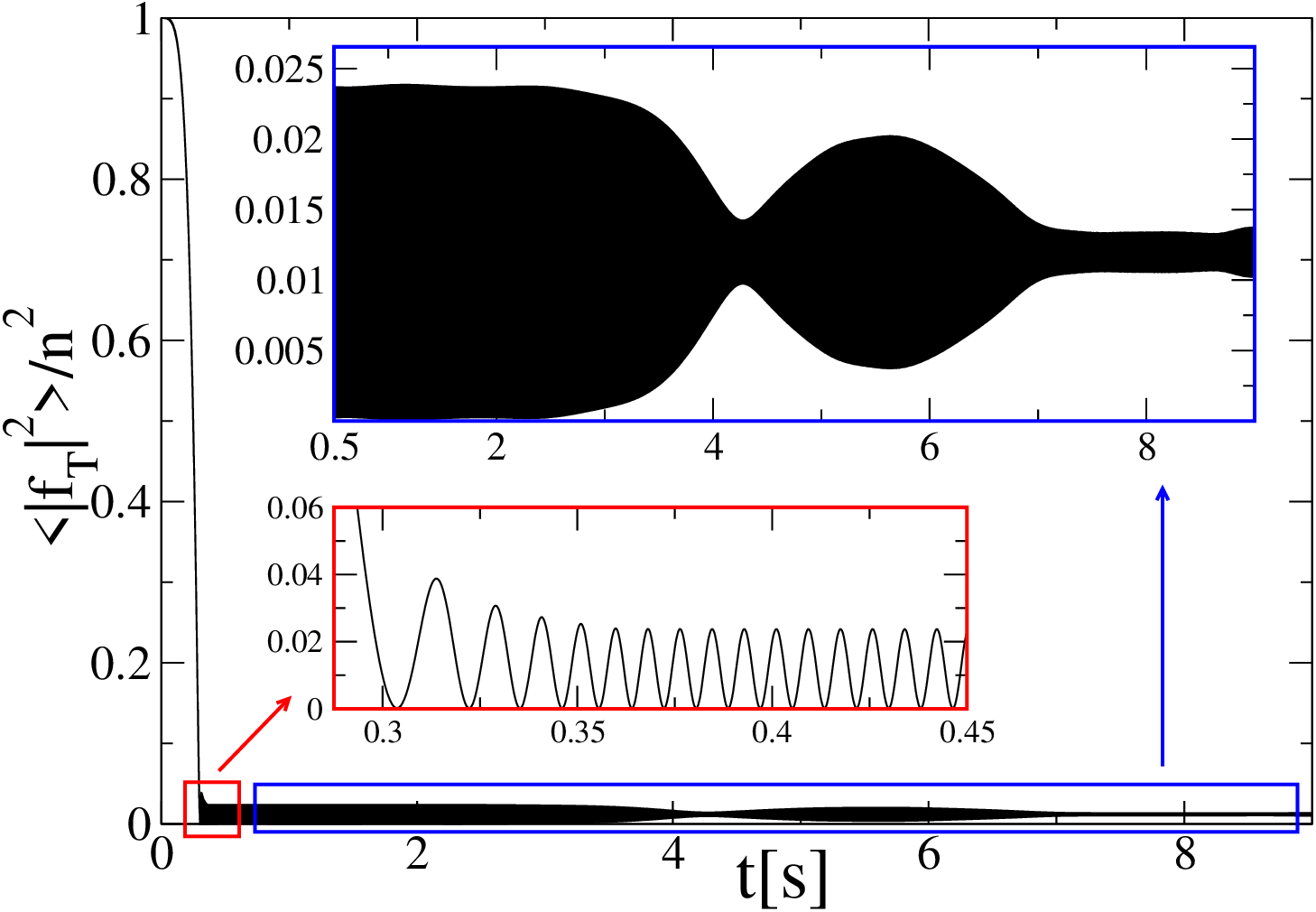}
\caption{(color) Dynamics of the 
         transverse magnetization in the inhomogeneous system during
         a quench with $\tau_Q=74ms$. The change of
         $q(t)$  stops at $3$ in the polar phase, 
	 $q(t\ge\sqrt{24}\tau_Q\approx0.36s)=3$, and $\delta=1\%$.
	 The insets show enlarged data from the main plot
	 corresponding to the regions depicted with boxes. 
	 The period of oscillations is so small that individual 
	 oscillations are not resolved in the main plot and the upper inset.
	 The lower inset does not show the SMA result as it is 
	 practically indistinguishable from the inhomogeneous calculation.
	 Arrows indicate which inset belongs 
	 to what box. 
	 The initial state for that evolution is the same as in Fig.
	 \ref{stop2_disorder}.
} 
\label{stop3_disorder}
\end{figure}

Now we would like to look at the system dynamics in the polar phase.
In Sec. \ref{sec4} we have observed  that the transverse magnetization 
undergoes 
periodic oscillations in a homogeneous problem after stop of driving. 
Now we see dephasing
dynamics spoiling this picture after a time inversely proportional 
to the disorder magnitude and the quench time $\tau_Q$. The latter observation 
results from the fact that the slower we go at fixed $\delta$, the more
significant the disorder is compared to the system magnetization in the polar
phase.

Fig. \ref{stop2_disorder} presents the evolution with $\tau_Q=74ms$
and $\delta=1\%$. The change of $q$ stops at the critical point there.
For these parameters the scaling exponents 
are very close to the Single Mode Approximation result
(see Table \ref{table}). As is illustrated in Fig. 
\ref{stop2_disorder}, however, dephasing takes place after about $7-9$
oscillations. To observe more magnetization oscillations 
closely following the SMA result, for twice that time,
$\delta$ has to be about one order of magnitude  smaller. 
Therefore, the system dynamics
is quite sensitive to inhomogeneities when the condensate is left at the
critical point. This should result from the coupling between different $k$ 
modes due to nonlinear terms in the evolution equation.

Fig. \ref{stop3_disorder} presents the evolution where driving
stops at $q=3$ (similarly as in Fig. \ref{dampl}). The parameters 
and the initial state taken for this simulation are the same as in 
 Fig. \ref{stop2_disorder}.
This time the undriven  dynamics takes place in the regime where 
nonlinear couplings are small. As a result, 
the dephasing occurs after hundreds of oscillations even though the disorder
amplitude, $\delta=1\%$, is the same as in Fig. \ref{stop2_disorder}.

To explain dephasing when the evolution stops away from a critical point, 
we can linearize the coupled Gross-Pitaevskii equations.
Below we do it in an explicitly 3D configuration to 
use the same notation as in Secs. \ref{sec3} and \ref{sec4} (the 1D result
is qualitatively the same). We assume that 
$$
\Psi = \sqrt{n}\left(
\begin{array}{c}
\delta\psi_1({\bf r},t) \\  1+\delta\psi_0({\bf r},t) \\
\delta\psi_{-1}({\bf r},t)
\end{array}
\right)\exp\left(-i\frac{\mu}{\hbar}t\right)
$$
where the chemical potential is $\mu=nc_0$, $|\delta\psi_m|\ll1$,
and $\int d{\bf r}(\delta\psi_0+\delta\psi_0^*)\equiv0$ to keep $\int d{\bf r}
\Psi^\dag\Psi=N+O(\delta\psi^2_m)$. In leading order 
$$
f_T = \sqrt{2}n(\delta\psi^*_1+\delta\psi_{-1}),
$$
and its dynamics at fixed $q$ is given by 
$$
\ddot f_T = -\frac{\hbar^2}{4M^2}\nabla^4f_T + (q-1)\frac{n|c_1|}{M}\nabla^2f_T
- \tau_0^{-2}q(q-2)f_T,
$$
where the last term is known from (\ref{xxx}). That can be solved by the
substitution $f_T\sim\cos({\bf kr}-\omega(k) t)$, giving
$$
\hbar\omega(k) = \sqrt{\epsilon_k^2 + \epsilon_k2(q-1)n|c_1|+ n^2|c_1|^2q(q-2)}, 
$$
where $\epsilon_k= \hbar^2k^2/2M$.
When disorder is present, different $k$ modes will be occupied and 
they will oscillate at different frequencies $\omega(k)$, which will result 
in the dephasing observed in Fig. \ref{stop3_disorder}. 
The expression for $\omega(k)$ 
shall not be used at $q=2$, where linearized theory fails as was shown within 
the SMA in Sec. \ref{sec4}.

\section{Summary}
\label{sec6}

We have analyzed  transitions from the broken-symmetry phase to the
polar phase occurring on the finite  timescale $\tau_Q$. Our focus was
on slow transitions, when the condensate
goes out of equilibrium near a critical point.
We have presented a simple  theory predicting 
that the evolution will cease to be adiabatic before reaching the polar phase
at the distance $\sim\tau_Q^{-2/3}$ from the critical point. 
This result was found 
to be in excellent agreement with numerics. Applying the basic assumptions
of the adiabatic-impulse approach \cite{dorner,bodzio}, which originates from the Kibble-Zurek
theory of nonequilibrium dynamics of classical phase transitions \cite{kibble,zurek}, 
we have explained why 
the transverse magnetization of the condensate 
driven to the critical point scales as $\tau_Q^{-2/3}$ as well. 
Subsequently we have analyzed how the 
latter  result can be experimentally extracted from 
the observation of magnetization oscillations. Three cases were considered.
First, we have studied what happens when the quench stops at the critical 
point and the system follows free evolution. It was shown  that 
the periodic oscillations appear  with the repetition period 
coupled to the amplitude of transverse magnetization oscillations. The repetition 
period was found to scale as $\tau_Q^{1/3}$, where the exponent was directly 
related to the scaling of the transverse magnetization at the critical 
point. That nonlinear dynamics was exactly described within the mean-field 
approach in terms of Jacobi elliptic functions. Second, 
we have studied what happens when the system is driven 
 into the polar phase, and then undergoes a free evolution.
In that case we have shown that its excitation at the critical 
point can be easily extracted from the scaling of the amplitude of the free 
magnetization oscillations given by $\tau_Q^{-1}$. 
Third, we have studied what happens when the
condensate is driven without any interruptions not only in the broken-symmetry phase,
but also in the polar phase. In this case the amplitude of  driven magnetization oscillations 
reaches a non-zero asymptotic value: the condensate magnetization
is described as 
an unusual harmonic oscillator with anti-friction term that perfectly 
cancels the squeezing induced by the increase of the harmonic oscillator
frequency.

All the above results were obtained within a translationally invariant 
Single Mode Approximation whose range of applicability was determined
by introducing a controlled disorder into an initial wave function.
We have found out that the amount of disorder has to be quite small
to have the homogeneous system predictions applicable.

Future extensions of this work will include studies of  quantum 
phase transitions dynamics
in a harmonically trapped ferromagnetic condensate
and investigations of
the role of quantum fluctuations on the condensate dynamics.

\section{Acknowledgments}

We gratefully acknowledge the support of the U.S. 
Department of Energy through the LANL/LDRD Program 
for this work.


\begin{thebibliography}{99}
\bibitem{jednostki} It is instructive here to note that after the rescaling
                    $\psi_m\to\sqrt{n}\psi_m$ and $t\to\tau_0t$, where $\tau_0$ is
		    defined in Eq. (\ref{qhat}),
		    we arrive at dimensionless description where none 
		    of the following quantities explicitly appears: $\hbar$, $c_1$, $n$, $N$.
		    Therefore, as long as the SMA is used, all the results 
		    are obtained without specifying  
		    the system-dependent parameters $c_1$, $n$, $N$.
\bibitem{liniowe} By choosing magnetic field strength $\sim\sqrt{t}$ 
one gets $q(t)=t/\tau_Q$. In this case,
however, the system
gets slightly excited right at the start of the evolution
as a result of a fairly abrupt switch of $q$ 
(e.g., $\dot q(0)\neq0$). As a result, the magnetization 
noticeably oscillates around the ground state prediction in the entire 
broken-symmetry phase for intermediate quench rates. 
The same effect shows up for (\ref{q_od_t}), but to  much smaller extent.
\end{thebibliography}
\end{document}